\documentclass[11pt,a4paper]{article}

\usepackage{jheppub}
\usepackage[T1]{fontenc}
\usepackage{xcolor}
\usepackage{graphicx}
\usepackage{bm}
\usepackage{amsmath,amsfonts,amssymb}
\usepackage{mathtools}
\usepackage{braket}
\usepackage[normalem]{ulem}
\usepackage{setspace} 
\usepackage{here}
 
\begin{document}


\title{Dai-Freed anomaly in the standard model and topological inflation}
\author[a,b]{Masahiro Kawasaki,}
\author[c,b]{Tsutomu T. Yanagida}
\affiliation[a]{ICRR, University of Tokyo, Kashiwa, 277-8582, Japan}
\affiliation[b]{Kavli IPMU (WPI), UTIAS, University of Tokyo, Kashiwa, 277-8583, Japan}
\affiliation[c]{Tsung-Dao Lee Institute and School of Physics and Astronomy, Shanghai Jiao Tong University, 520 Shengrong Road, Shanghai, 201210, China}
\abstract{%
When we impose discrete symmetries in the standard model we have Dai-Freed global anomalies. 
However, interestingly if we introduce three right-handed neutrinos we can have an anomaly-free discrete $Z_4$ gauge symmetry. 
This $Z_4$ symmetry should be spontaneously broken down to the $Z_2$ symmetry to generate the heavy Majorana masses for the right-handed neutrinos. 
We show that this symmetry breaking naturally generates topological inflation, which 
is consistent with the CMB observations at present and predicts a significant tensor mode with scalar-tensor ratio $r > 0.03$. 
The right-handed neutrinos play an important role in reheating processes.
The reheating temperature is as high as $\sim 10^8$~GeV, and non-thermal leptogenesis successfully takes place.
}

\emailAdd{kawasaki@icrr.u-tokyo.ac.jp}
\emailAdd{tsutomu.tyanagida@sjtu.edu.cn}

\maketitle

\section{Introduction}

Anomalies give strong constraints on models based on quantum field theories. If there are anomalies of gauge symmetries, they must be canceled out to have consistent gauge symmetries in quantum field theories. The anomalies are classified into two classes. One is traditional local anomalies \cite{Adler:1969gk,Bell:1969ts} and the other is anomalies of large gauge transformations. Witten's  $SU(2)$ global anomaly \cite{Witten:1982fp} is a famous example for the latter class. Both are well described by the Dai-Freed viewpoint \cite{Dai:1994kq,Witten:2015aba}, and we call them the Dai-Freed anomalies \cite{Yonekura:2016wuc}. The Dai-Freed anomaly-free conditions result in additional constraints on particle physics models since they require non-trivial conditions on massless chiral fermions.

It is known that the standard model is Dai-Freed anomaly free. However, if we impose additional discrete symmetries, it is not necessarily anomaly free and we should add new fermions to cancel
the anomalies in the SM model. In fact, if we impose a discrete $Z_4$ gauge symmetry in the SM, the theory has the Dai-Freed anomalies\footnote{
The discrete $Z_4$ symmetry has anomalies on a non-trivial gravitational background. For instance, when we consider the K3 curved surface we have 2 zero modes for one Weyl fermion on this surface.
Under the $Z_4$ transformation, the product of the 2 zero modes changes its sign. 
Thus, we need an even number of the Weyl fermions to make the fermion measure invariant of the $Z_4$ symmetry. 
Now we have 15 Weyl fermions in one family in the standard model and the $Z_4$ is anomalous in the standard model. However, if introduce one right-handed neutrino in each family the theory can be well defined with the $Z_4$ gauge symmetry.}
. 
However, the introduction of one right-handed neutrino for each generation cancels out the anomalies \cite{Garcia-Etxebarria:2018ajm}. 
Therefore, $Z_4$ is regarded as the origin of the presence of right-handed neutrinos.

The right-handed neutrinos should obtain Majorana masses. 
The large Majorana masses are very important for inducing the observed small masses for the active neutrinos via the seesaw mechanism \cite{Minkowski:1977sc, Yanagida:1979as,Yanagida:1979gs,Gell-Mann:1979vob} and for generating the baryon-number asymmetry in the universe \cite{Fukugita:1986hr}.
A scalar boson $\phi$ is needed to give Majorana masses for the right-handed neutrinos.
The discrete $Z_4$ symmetry breaks down to the discrete $Z_2$ by the condensation of the scalar $\phi$. 
 However, the spontaneous breaking of the discrete $Z_4$ gauge symmetry creates domain walls, which can be removed  away by inflation. 
 The simplest possibility is to identify the scalar $\phi$ itself with the inflaton.
 In this case, reheating of the universe takes place thanks to the coupling between the scalar $\phi$ and the right-handed neutrinos.

In this paper, we show that topological inflation naturally occurs in a large parameter region of the present model with the discrete $Z_4$ gauge symmetry. The topological inflation is very attractive since it is free from the initial state tuning problem \cite{Vilenkin:1994pv,Linde:1994hy}. 
In this model the right-handed neutrinos play an important role in the reheating processes as well as in generating the baryon-number asymmetry in the present universe.

This paper is organized as follows.
In Section~\ref{sec:discrete_Z4}, we briefly describe $Z_4$ symmetry and introduce the right-handed neutrinos.
Dynamics and observational implications of the topological inflation model are described in Section~\ref{sec:topological_inf}.
Section~\ref{sec:conslusion_discussion} is devoted to the conclusion and discussion of our results.

\section{Discrete \texorpdfstring{$Z_4$}{Z4} gauge symmetry in the standard model}
\label{sec:discrete_Z4}

We introduce a discrete $Z_4$ gauge symmetry in the standard model (SM). The charges are shown in Table~\ref{table:Z4_charge}. Here, we use the $SU(5)$ representations for the SM particles for simplicity of the notations, but we consider its subgroup $SU(3)\times SU(2)\times U(1)$ as a gauge group. It is stressed \cite{Garcia-Etxebarria:2018ajm} that the SM with the $Z_4$ is Dai-Freed anomaly free if we introduce one right-handed neutrino, $N_i$, for each generation $i=1,2,3$. Thus, we assume the SM with three right-handed neutrinos. The anomaly-free nature might be understood by embedding the $Z_4$ into the well-known $U(1)_{B-L}$ gauge group, but it is not necessarily. It might be amusing that three right-handed neutrinos are required by the cancellation of the Dai-Freed anomalies in the SM with the gauged $Z_4$ symmetry.

We add a SM gauge singlet boson $\phi$ to generate Majorana masses for the right-handed neutrinos, $N_i$. We consider a coupling of the $\phi$ to $N_i N_i$ as
\begin{equation}
    \label{eq:inflaton-Rneutrino}
    L=\frac{y_i}{2}\phi N_iN_i~ +~ h.c..,
\end{equation}
with $y_i$ coupling constants.
Here, the boson $\phi$ should have the $Z_4$ charge ${2}$ mod 4, since $N_i$ have the  $Z_4$ charge 1 (see Table 1). This new boson $\phi$ with a $Z_4$ charge 2 does not generate any Dai-Freed anomalies. 

\begin{table}[t]
    \centering
    \begin{tabular}{|c|ccccc|}
        \hline
               & $~5^*~$ & $~10~$   & $~N~$  & $~H~$  & $~\phi~$ \\
        \hline\hline
         $~Z_4~$ &  $1$  & $1$    & $1$  & $2$  & $2$   \\
        \hline 
    \end{tabular}
    \caption{$Z_4$ charges. $H$ is the Higgs doublet boson in the SM.}
    \label{table:Z4_charge}
\end{table}

\section{Topological inflation and leptogenesis}
\label{sec:topological_inf}

In this paper, we consider the scalar $\phi$ to be the inflaton. The scalar $\phi$ is neutral since it carries a $Z_4$ charge 2. The potential $V$ of the inflaton $\phi$ is then given by
\begin{equation}
    V= v^4 -2 g v^4\phi^2 + k v^4\phi^4+ ....,
\end{equation}
where the potential is invariant of the discrete $Z_4$ gauge symmetry, $g, k$ are coupling constants and $v$ is the energy scale of the potential.
Here and hereafter we use Planck units with the Planck mass $M_p = 2.4\times 10^{18}~\mathrm{GeV} = 1$.
The minimal nontrivial potential is given by
\begin{equation}
    \label{eq:potential}
    V=v^4(1-g\phi^2)^2.
\end{equation}
Here, we have chosen coupling so that the minimal of the potential has the vanishing vacuum energy. 
The $\phi$ obtains the vacuum expectation value $\langle\phi\rangle= \pm \sqrt{1/g}$ in the vacua, and the discrete $Z_4$ symmetry is spontaneously broken down to the $Z_2$ symmetry. 
In this paper we adopt this minimal potential to show how successfully the present model reproduces an inflationary universe consistent with the observations. 
We neglect possible higher order terms like $\phi^4, \phi^6, \ldots$ in the parenthesis of the potential~\eqref{eq:potential}. 
The extension to a more general potential will be studied in the future.

Suppose that the inflaton $\phi$ sits at the origin of the potential ($\phi=0$) in the early universe.
This is realized, for example, if the inflaton field acquires a thermal mass through the coupling~\eqref{eq:inflaton-Rneutrino}.
Then, when the energy density of the universe becomes $v^4$, $Z_4$ symmetry is spontaneously broken, and domain wall formation starts.
The width of the domain wall is given by $\langle \phi \rangle/\sqrt{V(0)} \simeq (\sqrt{g}v^2)^{-1}$~\cite{Vilenkin:1994pv} which is comparable to or larger than the Hubble horizon $H^{-1}\simeq v^{-2}$ for $g \lesssim \mathcal{O}(1)$.
Thus, inflation takes place in horizons contained inside the domain walls.
This type of inflation is called topological inflation.
Topological inflation is attractive because it is free from the initial value problem \cite{Linde:1994hy,Vilenkin:1994pv}. 
Moreover, since the initial horizon at $H\simeq v^2$, where the inflaton field is nearly homogeneous, becomes much larger than the present horizon by inflation, the domain wall problem does not exist\footnote{
The domain walls do not re-enter the horizon until the present when the total e-folds of inflation are larger than $\sim 60$.
Furthermore, if the present accelerated expansion continues, domain walls never re-enter the horizon in the future. 
}. 

Now we discuss the dynamics of topological inflation.
From the potential~\eqref{eq:potential} 
we obtain the slow-roll parameters $\epsilon$ and $\eta$ as
\begin{align}
    \epsilon & = \frac{1}{2}\left( \frac{V'}{V}\right)^2 
        = \frac{8g^2\phi^2}{1- 2 g \phi + g^2\phi^2} ,
        \\[0.5em]
    \eta & = \frac{V''}{V}
        = \frac{-4g+ 12g^2\phi^2}{1- 2 g \phi + g^2\phi^2}.
\end{align}
When $g \ll 1$, the slow-roll parameters $\epsilon$ and $\eta$ are smaller than $1$ for $\phi \lesssim \langle \phi \rangle$, and hence inflation lasts until the inflaton reaches nearly the minimum of the potential. 
Defining $\chi$ as $\phi-\langle \phi \rangle$, the potential is dominated by the quadratic term $\sim \chi^2$ around $\phi = \langle \phi\rangle$. 
Thus, inflation changes from the hilltop type for $\phi \ll 1$ to the chaotic one near $\phi\sim \langle \phi \rangle$.

With the use of $\chi$, the potential and the slow-roll parameters are rewritten as 
\begin{align}
    V & =  4gv^4 \chi^2 (1+g^{1/2}\chi/2)^2 
    \\[0.5em]
    \epsilon & = 
        \frac{2(1+g^{1/2} \chi)^2}{\chi^2(1+ g^{1/2}\chi/2)^2} ,
        \\[0.5em]
    \eta & =  \frac{2+ 6g^{1/2}\chi +3g\chi^2}{\chi^2(1+g^{1/2}\chi/2)^2}.
\end{align}
The field value at the end of inflation is then given by $\chi_f \simeq -\sqrt{2}$.
The e-fold number during inflation ($N = \ln(a(t_f)/a(t))$) is 
\begin{align}
    N \simeq \int^{\chi_f}_{\chi_N} 
         \frac{d\phi}{\sqrt{2\epsilon}}
         = \frac{\chi_N -\sqrt{2}}{4g^{1/2}}
         + \frac{\chi_N^2-2}{8}
         -\frac{1}{4g}\ln\left[\frac{1+g^{1/2}\chi_N}{1+\sqrt{2}g^{1/2}}\right],
\end{align}
from which we obtain $\chi_N$ as a function of $N$.
In Table~\ref{table:chi_Nefold} we show $\chi_{50}$, $\chi_{55}$ and $\chi_{60}$ for several values of $g$. 
\begin{table}[t]
    \centering
    \begin{tabular}{|c|ccccc|}
        \hline
          $~g~$ &  $0.001$  & $0.002$    & $0.003$   & $0.004$   & $0.005$\\
        \hline\hline
         $\chi_{50}$ &  $-14.11$  & $-13.05$    & $-12.32$  & $-11.73$  & $-11.21$ \\
         $\chi_{55}$ &  $-14.60$  & $-13.48$    & $-12.70$  & $-12.06$  & $-11.51$ \\
         $\chi_{60}$ &  $-15.07$  & $-13.89$    & $-13.06$  & $-12.37$  & $-11.77$ \\
        \hline 
    \end{tabular}
    \caption{$\chi_{N}$ for $N=50, 55$ and $60$.}
    \label{table:chi_Nefold}
\end{table}

\begin{figure}[t]
    \centering
    \includegraphics[width=.7\textwidth]{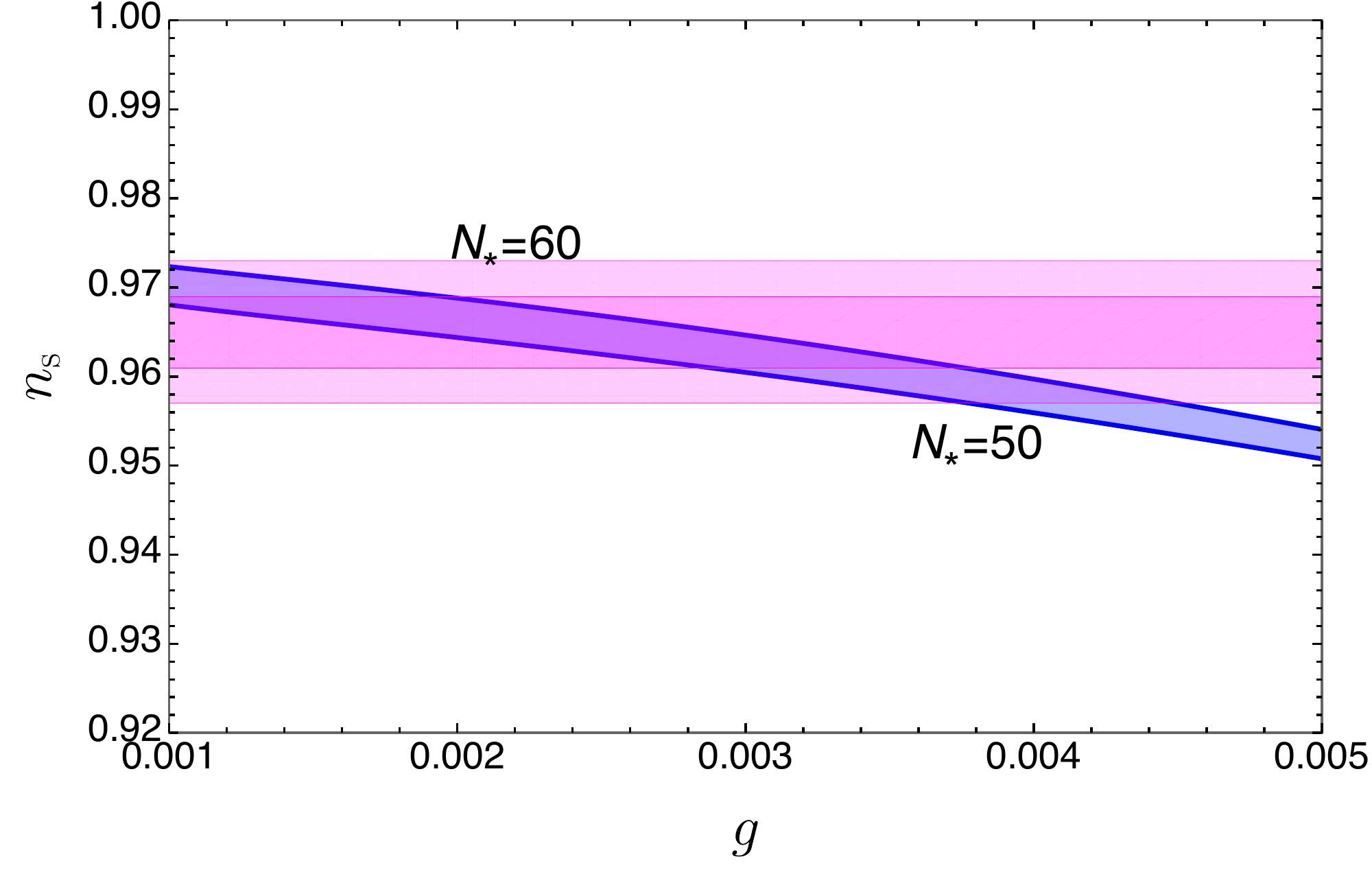}
    \caption{%
        Spectral index as a function of $g$ for $N_*=50-60$.
        The observed value with $1\sigma$ and $2\sigma$ errors is shown by the magenta-shaded region.
    }
    \label{fig:spec_index}
    \end{figure}
\begin{figure}[t]
    \centering
    \includegraphics[width=.7\textwidth]{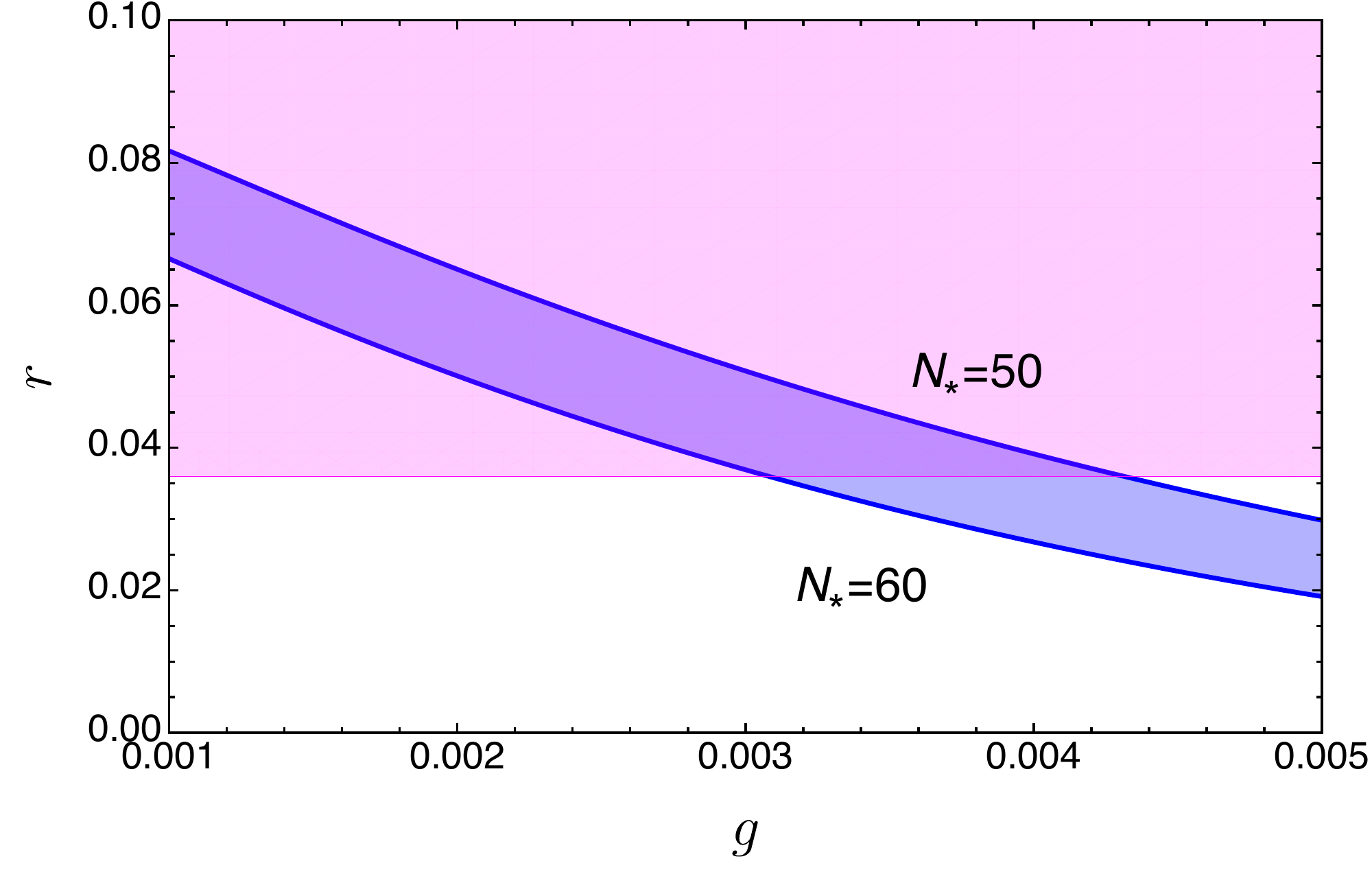}
    \caption{%
        Tensor-scalar ratio as a function of $g$ for $N_*=50-60$.
        The observed constraint $r < 0.036$ is shown by the magenta shaded region.
    }
    \label{fig:tensor_scalar}
    \end{figure}

\begin{figure}[t]
    \centering
    \includegraphics[width=.6\textwidth]{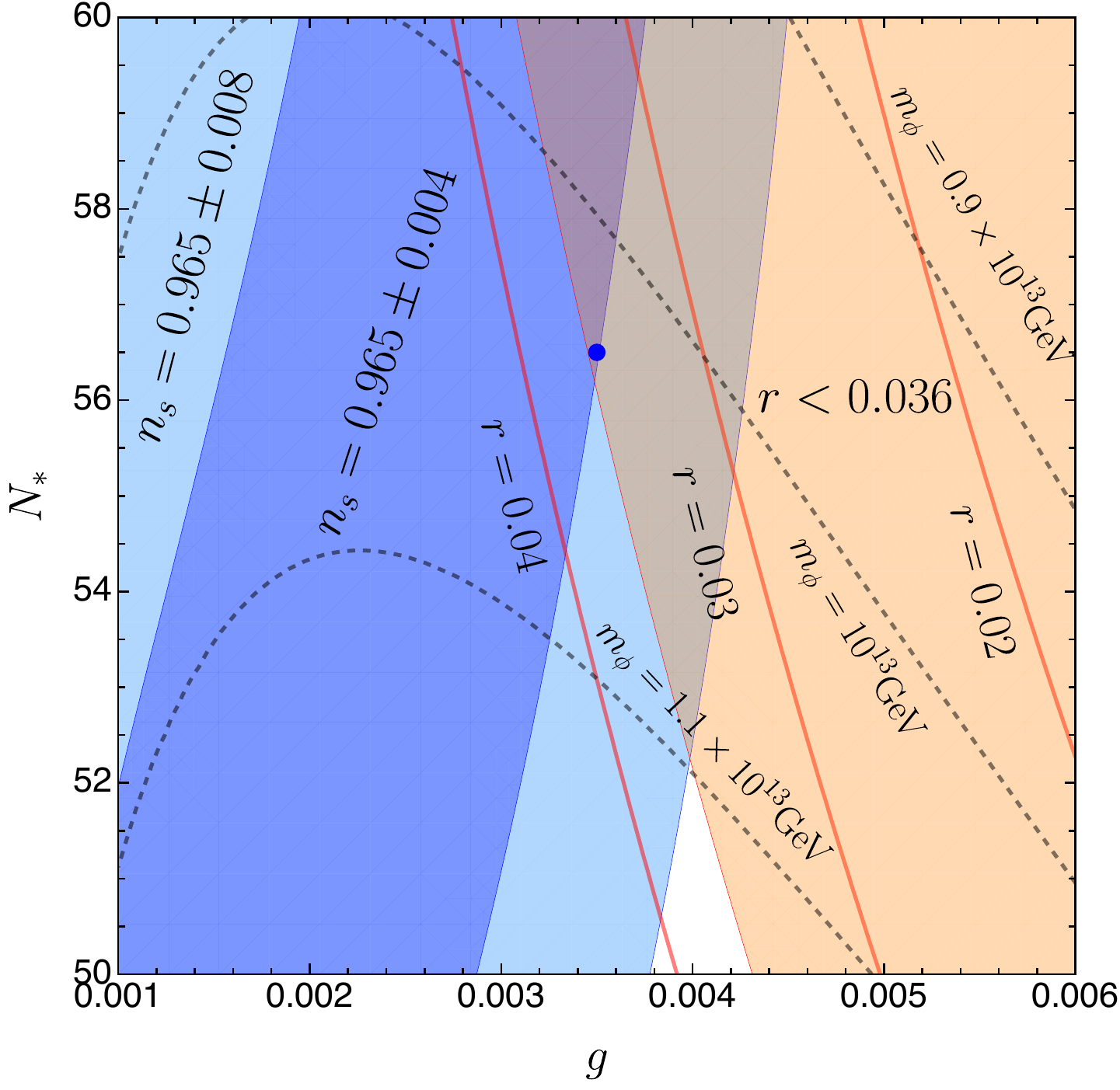}
    \caption{
        Allowed regions in the $g$--$N_*$ plane from the CMB observation.
        The (light) blue region is consistent with the observed spectral index within $1\sigma$ ($2\sigma$), and the orange region satisfies the constraint from the observation of the tensor-to-scalar ratio.
        The red lines denote $r= 0.02, 0.03$, and $0.04$.
        The dashed lines show the inflaton mass $m_\phi=(0.9, 1.0, 1.1)\times 10^{13}$~GeV.
        The blue point represents the reference parameters.}
    \label{fig:constraint}
    \end{figure}


The amplitude and spectral index of curvature perturbations at the CMB scale $k_* (= 0.002\mathrm{Mpc}^{-1})$ are written as
\begin{align}
    \label{eq:curvature_pert_amp}
    \mathcal{P}(k_*) & = \frac{V}{24\pi^2 \epsilon}
        = \frac{gv^4}{12\pi^2 }
           \frac{\chi_{N_*}^2(1+g^{1/2}\chi_{N_*}/2)^4}{(1+g^{1/2}\chi_{N_*})^2},\\[0.5em]
    \label{eq:curvature_pert_index}
    n_s & = 1-6\epsilon + 2\eta
        = 1- \frac{12(1+g^{1/2} \chi_{N_*})^2}
                    {\chi_{N_*}^2(1+ g^{1/2}\chi_{N_*}/2)^2}
            +\frac{4+ 12g^2\chi_{N_*} +6g\chi_{N_*}^2}
                    {\chi_{N_*}^2(1+g^{1/2}\chi_{N_*}/2)^2},
\end{align}
where $N_*$ is the e-fold number when the scale $k_*$ exits the Hubble horizon.
From Eq.~\eqref{eq:curvature_pert_amp} the inflation scale $v$ is  calculated as
\begin{align}
    v = \left[
        \frac{12\pi^2\mathcal{P}_\zeta(k_*)}{g}\,
        \frac{(1+g^{1/2}\chi_{N_*})^2}{\chi_{N_*}^2(1+g^{1/2}\chi_{N_*}/2)^4}
    \right]^{1/4}.
\end{align}
%

The model parameters $g$ and $v$ can determined by comparison with the Planck observation $\mathcal{P}_\zeta(k_*) = 2.1\times 10^{9}$ and $n_s= 0.965\pm 0.004\,(1\sigma)$~\cite{Planck:2018vyg}.
The predicted spectral index for $N_*= 50-60$ is shown in Fig.~\ref{fig:spec_index} where we also show the observed value with $1\sigma$ and $2\sigma$ errors.
The tensor-scalar ratio $r$ is given by $r= 16\epsilon$ and shown in Fig.~\ref{fig:tensor_scalar} together with the observational constraint $r< 0.036$~\cite{BICEP:2021xfz}.
The constraints from the observation of the spectral index and tensor-to-scalar ratio are shown in Fig.~\ref{fig:constraint} in the $g-N_*$ plane.
From the figure it is seen that the model is consistent with the observed $1\sigma$ ($2\sigma$) range of the spectral index and the upper bound on the tensor-to-scalar ratio if $0.003 \lesssim  g \lesssim  0.0036 (0.0045)$ for $N_* < 60$.

Hereafter we take $g=0.0035$ and $N_*=56.5$ as our reference values, which gives $n_s = 0.961$ and $r= 0.035$. 
The inflaton field value at $N=N_*$ is then calculated as
\begin{align}
    \chi_* =  -12.5M_\mathrm{p}.
\end{align}
The vacuum expectation value of the inflaton is given by $\langle \phi \rangle \simeq 16.9$.
Using the observed amplitude $\mathcal{P}_\zeta(k_*)$ the inflaton scale $v$ and the Hubble parameter during inflation are obtained as
\begin{align}
    v & \simeq 1.45\times 10^{16}~\mathrm{GeV} \\
    H_\mathrm{inf} & \simeq 4.67\times 10^{13}~\mathrm{GeV}.
\end{align}

Let us estimate the reheating temperature in this model.
The inflaton mass $m_\phi$ is 
\begin{equation}
    m_\phi = 2gv^2\langle \phi\rangle = 2\sqrt{g}v^2
    \simeq 1.03 \times 10^{13}~\mathrm{GeV}.
\end{equation}
Here notice that the inflaton mass is almost independent of $g$ as shown in Fig.~\ref{fig:constraint}.
This is because $v$ is proportional to $g^{-1/4}$ besides a weak dependence of $\chi_{N_*}$ on $g$.
The inflaton $\phi$ decays the right-handed neutrino through the interaction~\eqref{eq:inflaton-Rneutrino}
if a right-handed neutrino mass $m_{N_i}$ is less than $m_\phi/2$.
The decay rate is written as
\begin{align}
    \Gamma_\phi & = \frac{1}{8\pi} y_i^2  m_\phi 
    = \frac{1}{8\pi}\frac{m_{N_i}^2}{|\langle \phi \rangle|^2}m_\phi.
\end{align}
Here we have used  $m_{N_i} = y_i |\langle\phi\rangle|$.
From the decay rate we obtain the reheating temperature 
\begin{align}
    T_R & \simeq \left(\frac{\pi^2 g_*}{90}\right)^{_1/4}
    \Gamma_\phi^{1/2}
    \nonumber \\
    \label{eq:reheating_temp}
    & = 6.7\times 10^7~\mathrm{GeV}
       \left(\frac{m_{N_i}}{5\times 10^{12}\mathrm{GeV}}\right)
       \left(\frac{g_*}{100}\right)^{-1/4},
\end{align}
where $g_*$ is the relativistic degree of freedom. 
The reheating temperature is not high enough for thermal leptogenesis, but non-thermal leptogenesis is possible~\cite{Buchmuller:2005eh,Lazarides:1990huy,Kumekawa:1994gx,Asaka:1999yd,Asaka:1999jb,Giudice:1999fb}~\footnote{
We have suppressed an inflaton coupling $\phi^2 H^{\dagger}H$ since otherwise it generates too large mass for the Higgs boson. 
However, we may have this term if the mass term $m^2H^{\dagger}H$ cancels the induced Higgs boson mass. 
In this case, we have a fast $\phi$ decay to a pair of the Higgs bosons and the reheating temperature becomes much higher and the thermal leptogenesis works.
}.
When the inflaton mainly decays into the lightest right-handed neutrino $N_1$, the generated lepton number entropy ratio $n_L/s$ is 
\begin{align}
    \frac{n_L}{s} = - 3\times 10^{-10} 
        \left(\frac{T_R}{10^6\mathrm{GeV}}\right) 
        \left(\frac{m_{N_1}}{m_\phi}\right) 
        \left(\frac{m_3}{0.05\mathrm{eV}}\right) \delta_\mathrm{eff},
\end{align}
where $m_3$ is the heaviest neutrino mass and $\delta_\mathrm{eff}$ is the degree of CP violation ($\delta_\mathrm{eff} \le 1$).
The lepton asymmetry is converted into the baryon asymmetry given by $n_B/s = -(8/23)\,n_L/s$.
With the reheating temperature~\eqref{eq:reheating_temp} we have a successful non-thermal leptogenesis.

\section{Conclusion and Discussion}
\label{sec:conslusion_discussion}
If we impose a discrete $Z_4$ gauge symmetry in the standard model (SM), we need extra chiral fermions to cancel the Dai-Freed global anomalies. 
The most simple candidate of the extra fermions is three right-handed neutrinos. 
This might be the reason why we have right-handed neutrinos. This discrete $Z_4$ symmetry must be broken down to $Z_2$ by a scalar $\phi$ condensation to generate Majorana masses for the right-handed neutrinos. 
However, the spontaneous breaking of the discrete symmetry causes too many domain walls in the early universe. 
The most simple solution to this problem is to identify the scalar $\phi$ with the inflaton, generating topological inflation.

In this paper, we have shown that the above topological inflation is consistent with the present observations. Furthermore, we have shown that the tensor-scalar ratio $r$ is predicted as $r>0.03$ using the scalar mode observation $n_s=0.095\pm 0.008\, (2\sigma)$~\footnote{
The supersymmetry extension of the present model is also Dai-Freed anomaly free. 
In this extension the inflaton potential becomes identical to the potential in \cite{Choi:2022ssv} and the constraint on the scalar-tensor ratio $r$ is much weaker as shown in \cite{Choi:2022fce}.
}. 
This prediction on the tensor mode will be tested in near future experiments like Simons Observatory~\cite{SimonsObservatory:2018koc} and LiteBIRD~\cite{LiteBIRD:2022cnt}.

In the present model, reheating takes place through the inflaton coupling to the lightest right-handed neutrino $N_1$. 
We have found that the reheating temperature $T_R \simeq 10^{8}$ GeV and we have a successful non-thermal leptogenesis for the baryon asymmetry in the present universe.

In general, topological inflation is eternal because the quantum fluctuation $\delta_q\phi\simeq H_\mathrm{inf}/2\pi$ dominates over the classical displacement $\delta_c \phi \simeq \dot{\phi}H_\mathrm{inf}^{-1}$ near origin of the potential.
Using the potential~\eqref{eq:potential} the condition for eternal inflation is written as
\begin{equation}
    \phi \lesssim \phi_\mathrm{eternal} =\frac{3v^2}{8\pi g} 
    = \frac{3v^2}{8\pi g^{1/2}}\langle\phi\rangle.
\end{equation}
For $g\sim \mathcal{O}(10^{-3})$, we found $v \sim 10^{-2}M_p$ and $\phi_{N_*} \gtrsim 0.1\langle \phi\rangle$.
Therefore, $\phi_{N_*} \gg \phi_\mathrm{eternal}$ and our analysis based on classical slow-roll approximation is valid.


 \section*{Acknowledgements}

T. T. Y. thanks Kazuya Yonekura for the discussion on the Dai-Freed anomaly. T. T. Y. is supported by the China Grant for
Talent Scientific Start-Up Project and by Natural Science
Foundation of China (NSFC) under grant No.\,12175134,
JSPS Grant-in-Aid for Scientific Research
Grants No.\,19H05810, 
and World Premier International Research Center
Initiative (WPI Initiative), MEXT, Japan.
M. K. is supported by JSPS KAKENHI Grant Nos. 20H05851(M.K.) and 21K03567(M.K.). 

\bibliographystyle{JHEP}   
\bibliography{Ref}

\end{document}